\documentclass[9pt]{article}
\usepackage{graphicx}

\newcommand\pubnumber{DPF2013-219}
\newcommand\pubdate{\today}

\def\brown{Department of Physics\\
Brown University, Providence, Rhode Island 02912}

\def\Title#1{\begin{center} {\Large #1 } \end{center}}
\def\Author#1{\begin{center}{ \sc #1} \end{center}}
\def\Address#1{\begin{center}{ \it #1} \end{center}}

\newcommand\pubblock{\rightline{\begin{tabular}{l} \pubnumber\\
         \pubdate  \end{tabular}}}
\newenvironment{Abstract}{\begin{quotation}  }{\end{quotation}}
\newenvironment{Presented}{\begin{quotation} \begin{center} 
             PRESENTED AT\end{center}\bigskip 
      \begin{center}\begin{large}}{\end{large}\end{center} \end{quotation}}

\def\beq{\begin{equation}}
\def\eeq#1{\label{#1}\end{equation}}
\def\eeqn{\end{equation}}

\def\beqa{\begin{eqnarray}}
\def\eeqa#1{\label{#1}\end{eqnarray}}
\def\eeqan{\end{eqnarray}}

\let\bar=\overbar

\def\Dslash{\not{\hbox{\kern-4pt $D$}}}
\def\dslash{\not{\hbox{\kern-2pt $\del$}}}

\def\msb{{\bar{\ssstyle M \kern -1pt S}}}

\begin{document}
\begin{titlepage}
\pubblock

\vfill
\Title{Search for Exotic Top Partners at $\sqrt{s}$ = 8 TeV}
\vfill
\Author{Saptaparna Bhattacharya, on behalf of the CMS Collaboration}
\Address{\brown}
\vfill
\begin{Abstract}
We present searches for heavy top and bottom quark partners at CMS using data collected at $\sqrt s$ = 8 TeV. Such partners, if vector-like, occur in models such as the Little Higgs and Large Extra Dimensions. Fermionic top partners could also occur in composite Higgs models. The searches presented here span a wide range of final states, from lepton plus jets to multi-leptonic, and exclusion limits are set on mass and production cross sections as a function of branching ratios of the heavy quarks to their decay products.
\end{Abstract}
\vfill
\begin{Presented}
DPF 2013\\
The Meeting of the American Physical Society\\
Division of Particles and Fields\\
Santa Cruz, California, August 13--17, 2013\\
\end{Presented}
\vfill
\end{titlepage}

\section{Introduction}

Many theoretical extensions that predict physics beyond the standard model (BSM) propose the existence of fermionic partners of the top-quark. These quarks could either be vector-like or could occur in composite-Higgs models (T$_{5/3}$). The search for vector-like T quarks includes T decaying to bW, tZ and tH. These final states are interesting since the newly discovered Higgs boson is used as a probe for new physics (Sec.~\ref{sec:Tquark}). Composite Higgs models that posit the existence of T$_{5/3}$ are not ruled out by the presence of a SM Higgs at 125 GeV, since T$_{5/3}$ pair production does not contribute significantly to the Higgs cross section (Sec.~\ref{sec:T53}). Searches for models with bottom quark partners, b$^{\prime}$, have also been carried out at CMS~\cite{cms_exp}. The reinterpretation of an R-parity violating (RPV) SUSY search as a b$^{\prime}$ search is presented in Sec.~\ref{sec:SUSY}.

\section{Search for vector-like top partners ~\cite{cms_search1}}\label{sec:Tquark}

Vector-like quarks appear in many BSM scenarios  such as the Little Higgs models or models with Extra Dimensions or MSSM. A vector-like quark, T is pair produced via gluon fusion and quark-anti-quark annihilation and decays to bW, tH and tZ. This leads to busy final states with multiple bosons and b-jets.

\subsection{The Single Lepton Channel}\label{sec:Singlelep}

The single lepton channel requires an isolated lepton with p$_{T} >$ 32 GeV and $|\eta| <$ 2.1. The event selection requires a minimum transverse momentum (p$_{T}$) threshold (35 GeV) on the first four leading (anti-k$_{T}$ with radius parameter, $R$, 0.5 (AK5)) jets~\cite{ak51}. Jet substructure variables are used to tag ÒtopÓ and ÒWÓ jets. W-tagging uses a jet-pruning algorithm which takes Cambridge-Aachen (CA) jets of distance parameter, R, of 0.8 as inputs. The jets from highly boosted top quarks are merged into one jet using a top-tagging algorithm.

\begin{figure}[!h]
\includegraphics[width=0.48\textwidth]{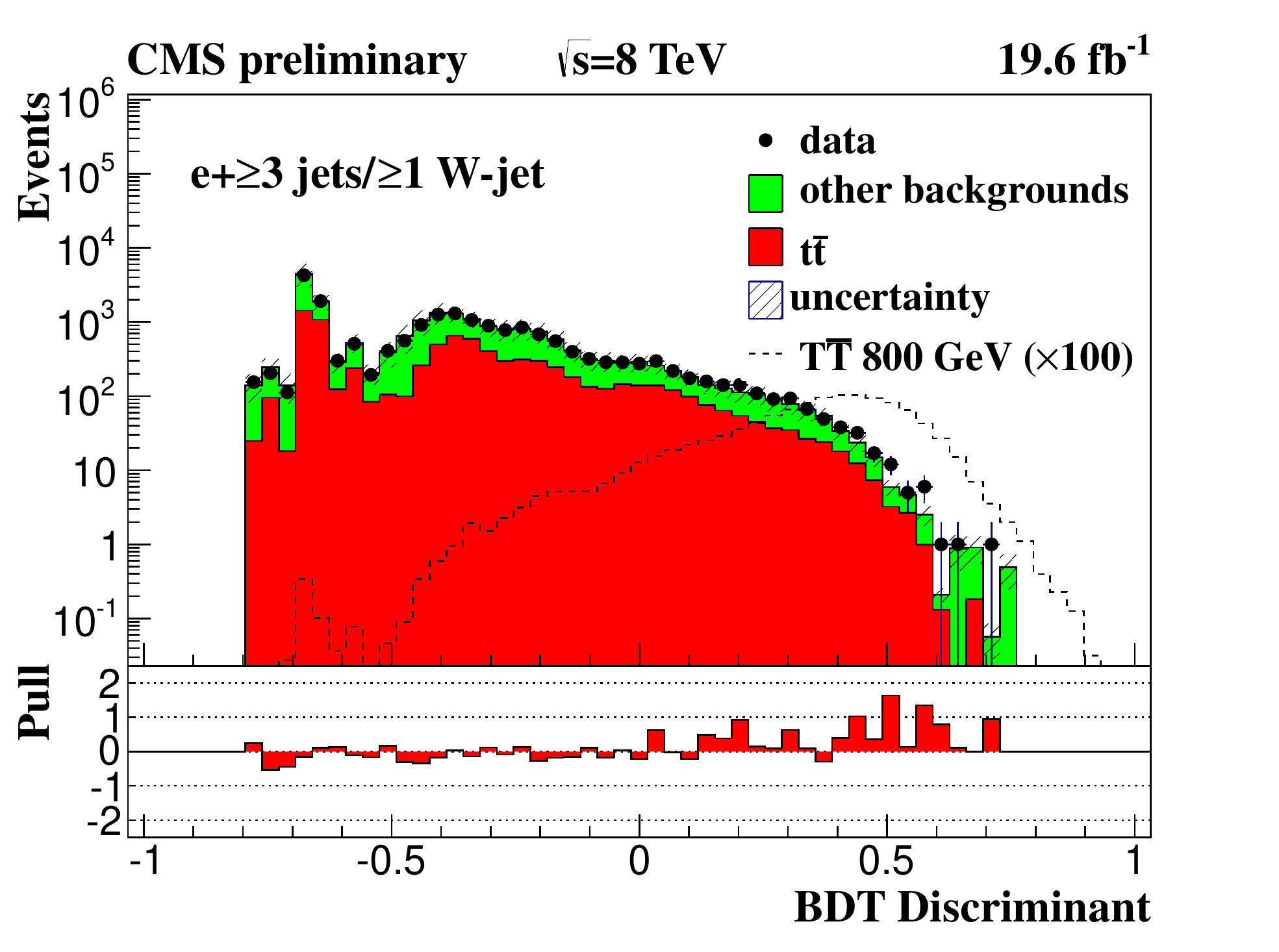}
\includegraphics[width=0.48\textwidth]{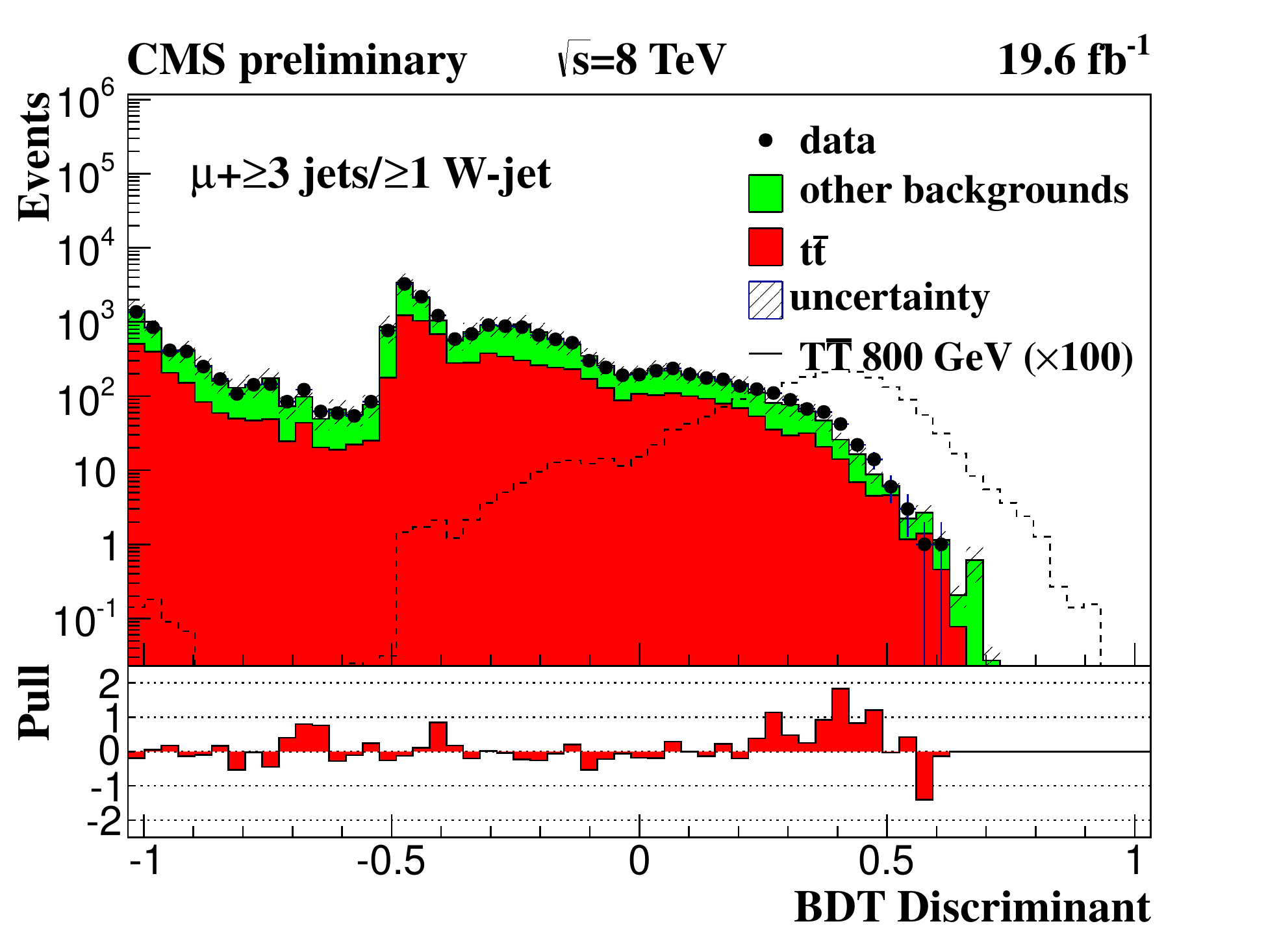}
\caption{BDT distributions in events with 1 W-jet and at least 3 AK5 jets.\label{fig:BDT1}}
\end{figure}

\begin{figure}[!h]
\includegraphics[width=0.48\textwidth]{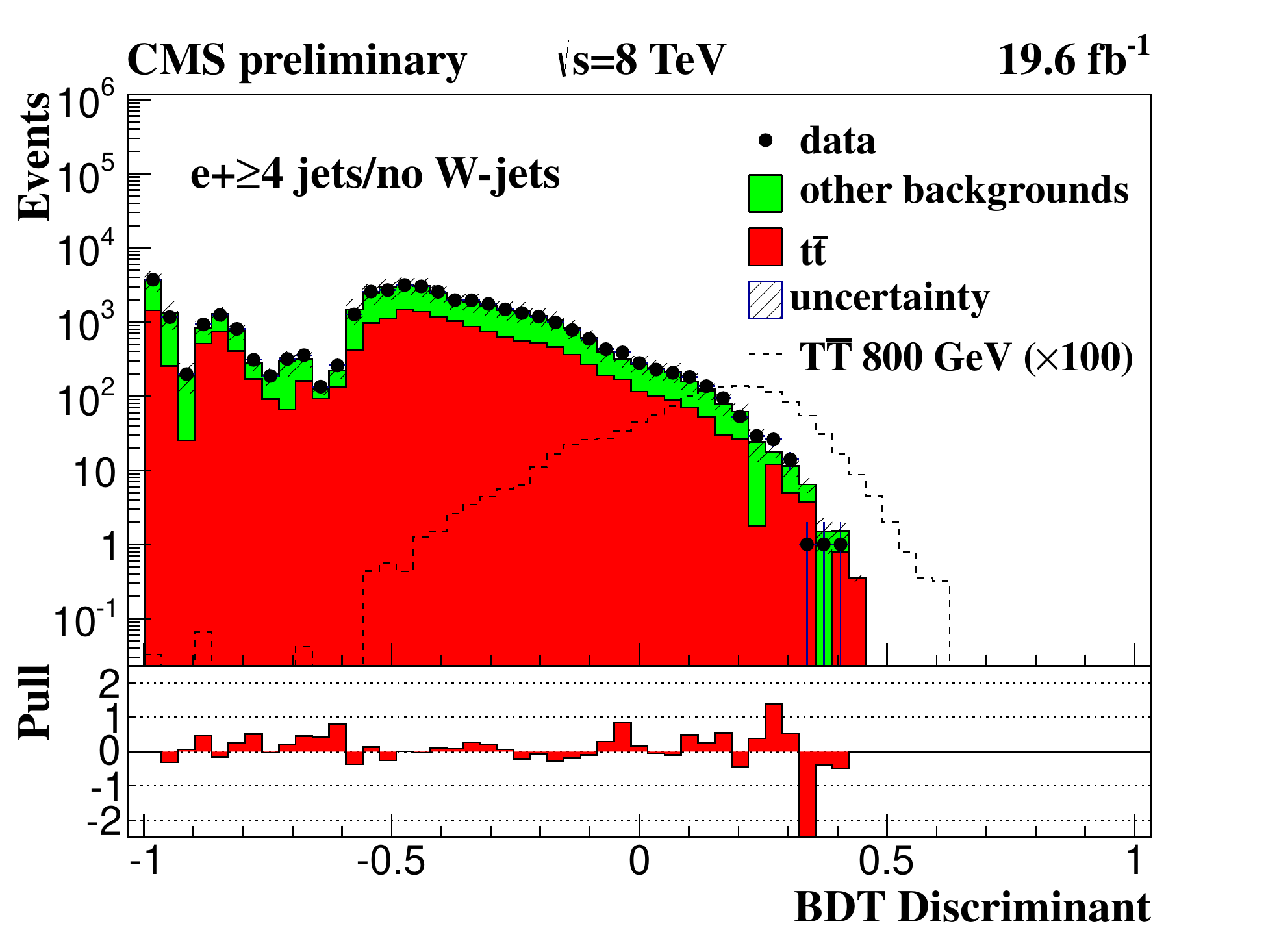}
\includegraphics[width=0.48\textwidth]{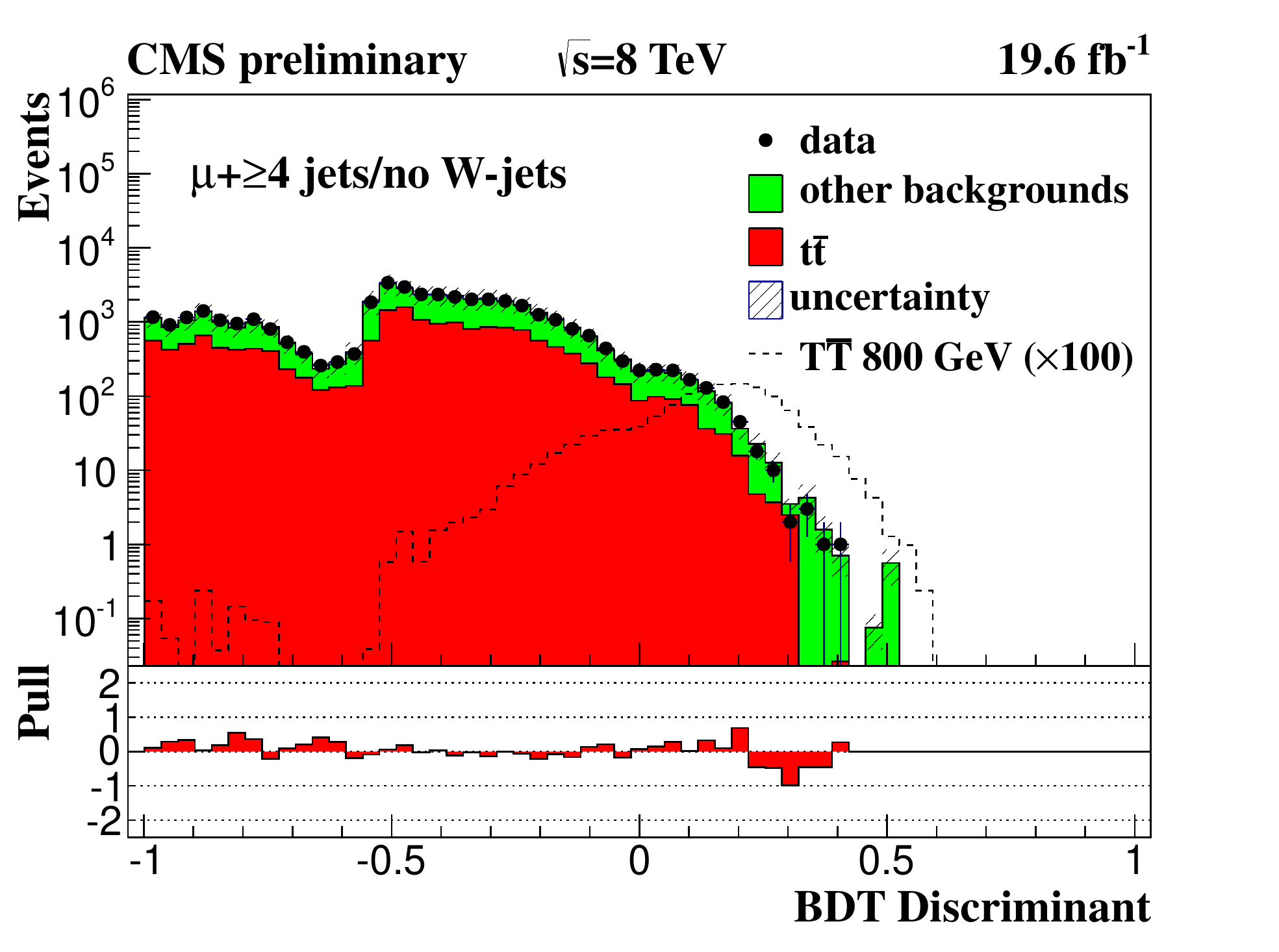}
\caption{BDT distributions in events with 1 W-jet and at least 4 AK5 jets with p$_{T} > $ 35 GeV. \label{fig:BDT2}}
\end{figure}

A boosted decision tree (BDT) is used to obtain optimal separation between signal and SM background ($t\bar{t}$, W and Z boson production processes constitute 96\% of the total background). Two separate event categories are constructed based on the presence (or absence) of a W tagged jet. The input variables to the BDT are the following kinematic variables: jet multiplicity, b-jet multiplicity, the sum of the transverse momenta of the selected jets (H$_{T}$),  missing transverse momenta (E$_{T}$),  p$_{T}$ of the leading lepton, p$_{T}$ of the 3rd and the 4th jets. In the W tagged events, the p$_{T}$ of the W-jets and the number of top-tagged jets are used. The BDT discriminants obtained are shown in Fig.~\ref{fig:BDT1}, \ref{fig:BDT2}. The event yields are quoted in Table~\ref{tab:yields1}. 

\begin{table}[htbp]
\begin{center}
\caption{The total number of expected and observed events in the single-lepton channel.}\label{tab:yields1}

\begin{tabular}{lr@{$\pm$}lr@{$\pm$}l}
\hline
lepton flavor           &\multicolumn{2}{c}{muon} & \multicolumn{2}{c}{electron} \\
\hline
$t{\bar{t}}$                & 36700                 & 5500              & 35900     & 5400 \\
single top              & 2190                  & 1101           & 2100         & 1000 \\
W                       & 19200                 & 9700         & 18200  & 9200 \\
Z                       & 2170          & 1100              & 2000      & 1000 \\
multijets               & \multicolumn{1}{r}{0}& & 1680               & 620\\
$t{\bar{t}}$  W                & 144           & 72                & 137               & 68 \\
$t{\bar{t}}$ Z                & 109           & 54                & 108               & 54 \\
$t{\bar{t}}$  H                & 570               & 280          & 570                & 285 \\
WW/WZ/ZZ        & 410           & 205          & 400            & 200 \\
\hline
total background & 61500                & 13700        & 61100  & 13500 \\
data            & \multicolumn{1}{r}{58478}  &                & \multicolumn{1}{r}{57743}  \\
\hline\hline
\end{tabular}
\end{center}
\end{table}

\subsection{The Multilepton Channels}\label{sec:Multilep}

\begin{figure}[!h]
\includegraphics[width=0.48\textwidth]{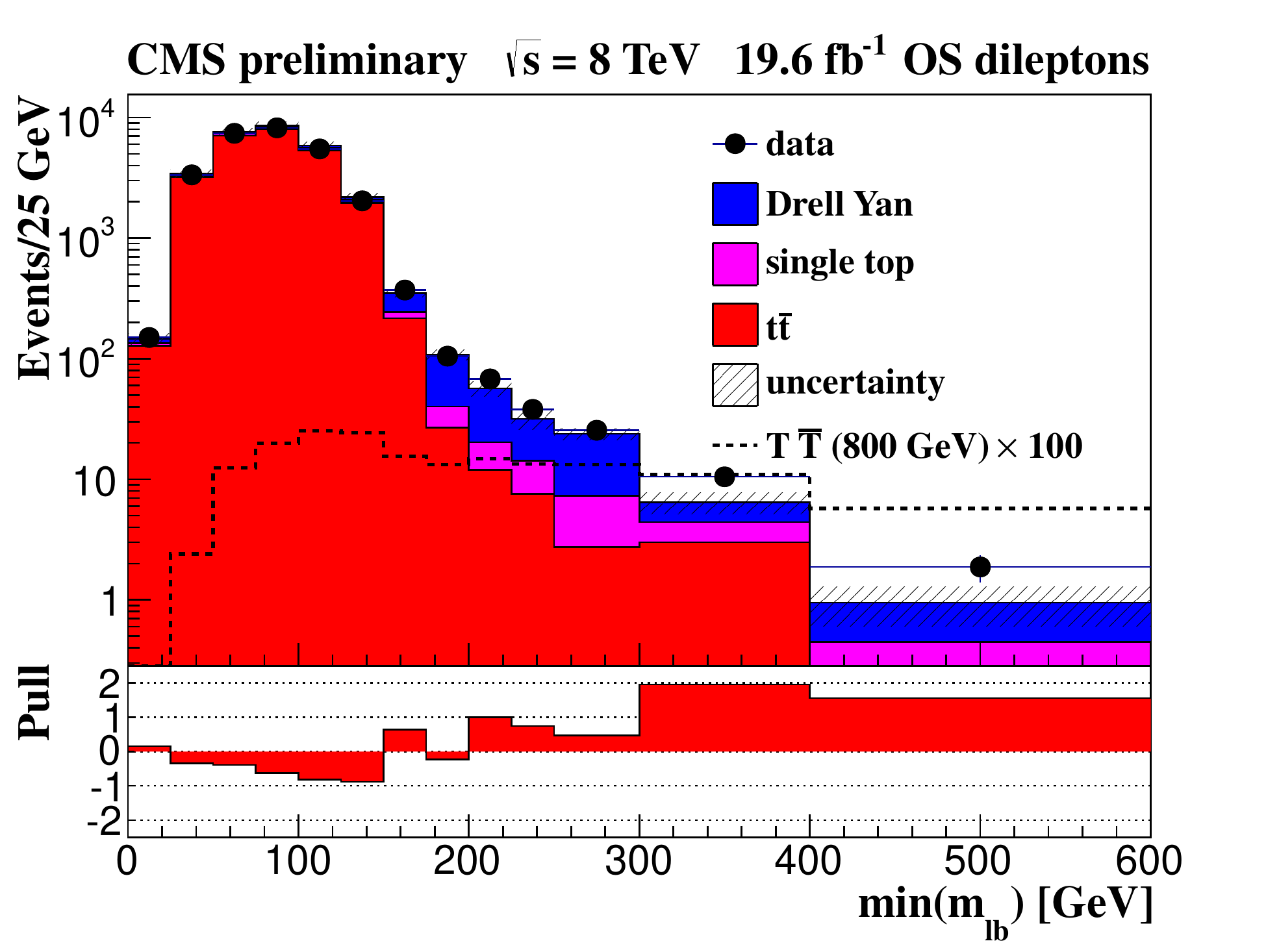}
\includegraphics[width=0.48\textwidth]{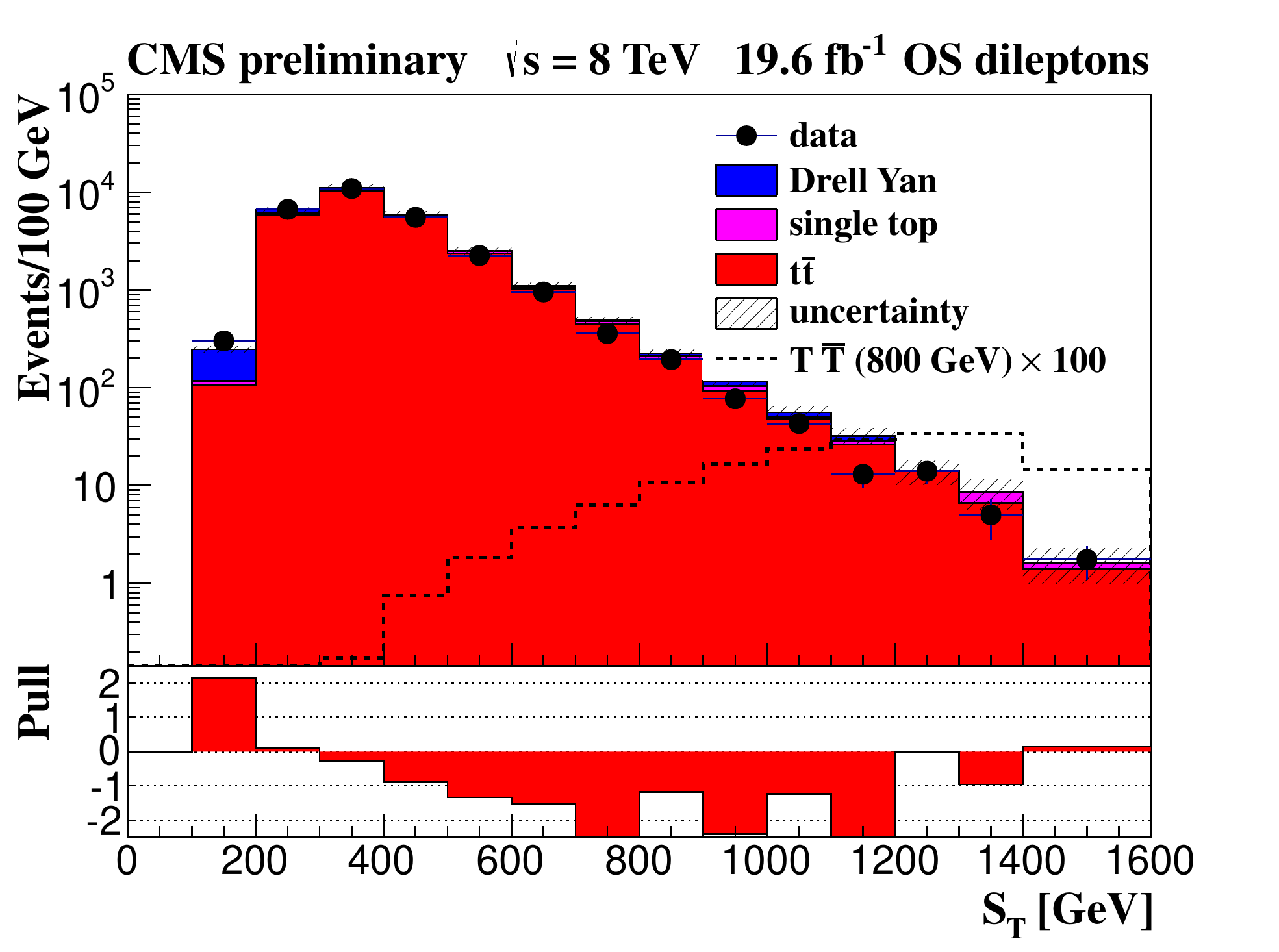}
\caption{Distribution of min(M$_{lb}$) and S$_{T}$ in the OS23 category\label{fig:Multilepton1}}
\end{figure}

In the multilepton channel, various event categories are constructed to be sensitive to a variety of leptonic final states. To this end, we create three broad categories: 

\begin{itemize}

\item Opposite sign (OS) lepton final state: In this category, we require two isolated leptons with opposite charge. The major irreducible backgrounds in this category are $t\bar{t}$ and Drell-Yan. While the Drell-Yan background can be minimized with the requirement of b-jets, $t\bar{t}$ completely mimics the TT $\rightarrow$ bWbW signal. Hence, we further subdivide the OS category based on requirements on the number of AK5 jets, b-jets, the minimum invariant mass of the lepton and b-jet (minM$_{lb}$, shown in Fig.~\ref{fig:Multilepton1}), the sum of the transverse momenta of all jets (H$_{T}$) and the sum of the transverse momenta of all jets, leptons and missing E$_{T}$ (S$_{T}$, shown in Fig.~\ref{fig:Multilepton1}).

\begin{itemize}

\item OS23: This category is constructed to be sensitive to the TT $\rightarrow$ bWbW signal. We require 2 or 3 jets, a Z-veto (since there are no real Z's in the signal), 1 b-jet, missing $E_{T}  >$ 30 GeV, H$_{T} >$ 300 GeV,  S$_{T} >$ 900 GeV and minM$_{lb} >$ 170 GeV. Since minM$_{lb}$ is directly proportional to the mass of the T quark, the last selection facilitates in the reduction of the $t\bar{t}$ background. 

\item OS5+: This category is sensitive to final states where the T quark decays to tH and tZ. We require at least 5 jets, 2 b-jets, missing  $E_{T}  >$ 30 GeV, H$_{T} >$ 500 GeV and S$_{T} >$ 1000 GeV.

\end{itemize}   

\item Same signed (SS) lepton final state: The backgrounds in this relatively clean channel are from SM processes like diboson and triboson decays. In addition to these backgrounds, there is a non-trivial contribution from fake leptons, either from events where both leptons are fake (from jets faking as leptons from multijet processes) or from events where one lepton is fake and the other prompt (from $t\bar{t}$). In the SS category, we find that fake leptons constitute 50\% of the total background. The final selection criteria involves requirements on the number of jets ($\geq$ 3 jets, 1 b-jet) , missing $E_{T}$ ($>$ 30 GeV), H$_{T}$ ($>$ 500 GeV) and S$_{T}$ ($>$ 700 GeV).

\item Trilepton final state: The backgrounds are SM processes like diboson and triboson decays. The fraction of fake leptons contributing to the total background is about 33\% in this category. A requirement on the number of jet (require $\geq$ 3 jets, 1 b-jet) , missing $E_{T} $($>$ 30 GeV), H$_{T}$ ( $>$ 500 GeV) and S$_{T}$ ( $>$ 700 GeV) is made.

 \end{itemize}

In both SS and trilepton categories, the instrumental backgrounds (fake leptons in both SS and trilepton category and charge misidentification in the SS category) are estimated directly from data. 
The final yields after the application of all event selection requirements are in Table~\ref{tab:yields_ll}

\begin{table}[htbp]
\begin{center}
\caption{Number of expected and observed events in the multilepton channel}\label{tab:yields_ll}

\begin{tabular}{lr@{}l@{$\pm$}lr@{}l@{$\pm$}lr@{}l@{$\pm$}lr@{}l@{$\pm$}l}
\hline
channel                 &\multicolumn{3}{c}{OS1} & \multicolumn{3}{c}{OS2} & \multicolumn{3}{c}{SS}     & \multicolumn{3}{c}{trileptons}\\
\hline
$t{\bar{t}}$                &  5&.2&1.9                     & 80& &12                       & \multicolumn{3}{c}{-}         & \multicolumn{3}{c}{-} \\
single top              &  2&.5&1.3                     & 2&.0&1.0                      & \multicolumn{3}{c}{-}         & \multicolumn{3}{c}{-} \\
Z                       &  9&.7&2.9                     & 2&.5&1.9                      & \multicolumn{3}{c}{-}         & \multicolumn{3}{c}{-} \\
$t{\bar{t}}$W               &  \multicolumn{3}{c}{-}        & \multicolumn{3}{c}{-}         &  5&.8&1.9                     & 0&.25&0.11    \\
$t{\bar{t}}$Z               &  \multicolumn{3}{c}{-}        & \multicolumn{3}{c}{-}         &  1&.83&0.93           & 1&.84&0.94  \\
WW              &  \multicolumn{3}{c}{-}        & \multicolumn{3}{c}{-}         &  0&.53&0.29           & \multicolumn{3}{c}{-} \\
WZ                      &  \multicolumn{3}{c}{-}        & \multicolumn{3}{c}{-}         & 0&.34&0.08                    & 0&.40&0.21  \\
ZZ                      &  \multicolumn{3}{c}{-}        & \multicolumn{3}{c}{-}         &  0&.03&0.00           & 0&.07&0.01  \\
WWW/WWZ/ZZZ/WZZ         &  \multicolumn{3}{c}{-}        & \multicolumn{3}{c}{-} &  0&.13&0.07  & 0&.08&0.04  \\
$t{\bar{t}}$WW      &  \multicolumn{3}{c}{-}        & \multicolumn{3}{c}{-}         &  \multicolumn{3}{c}{-}        & 0&.05&0.03  \\
charge mis-ID   &  \multicolumn{3}{c}{-}        & \multicolumn{3}{c}{-}         &  0&.01&0.00           & \multicolumn{3}{c}{-} \\
non-prompt      &  \multicolumn{3}{c}{-}        & \multicolumn{3}{c}{-}         &  7&.9&4.3                     & 0&.99&0.90  \\
\hline
total background & 17&.4&3.7            & 84& &12                       & 16&.5&4.8                     & 3&.7&1.3 \\
data                    &  20& \multicolumn{2}{c}{} & 86& \multicolumn{2}{c}{} & 18& \multicolumn{2}{c}{} & 2& \multicolumn{2}{c}{} \\
\hline\hline
\end{tabular}
\end{center}
\end{table}

\begin{figure}[!h]
\includegraphics[width=0.48\textwidth]{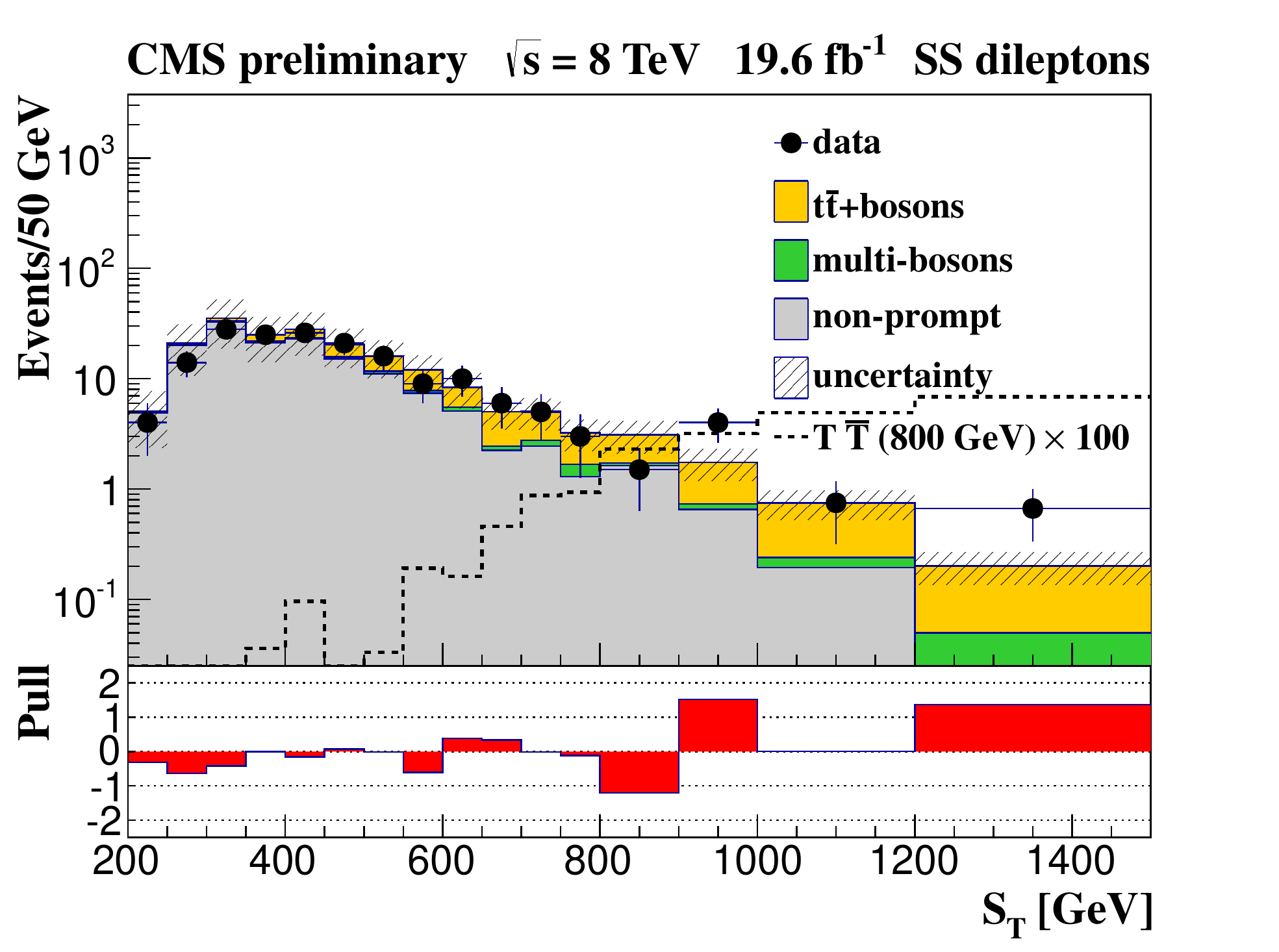}
\includegraphics[width=0.48\textwidth]{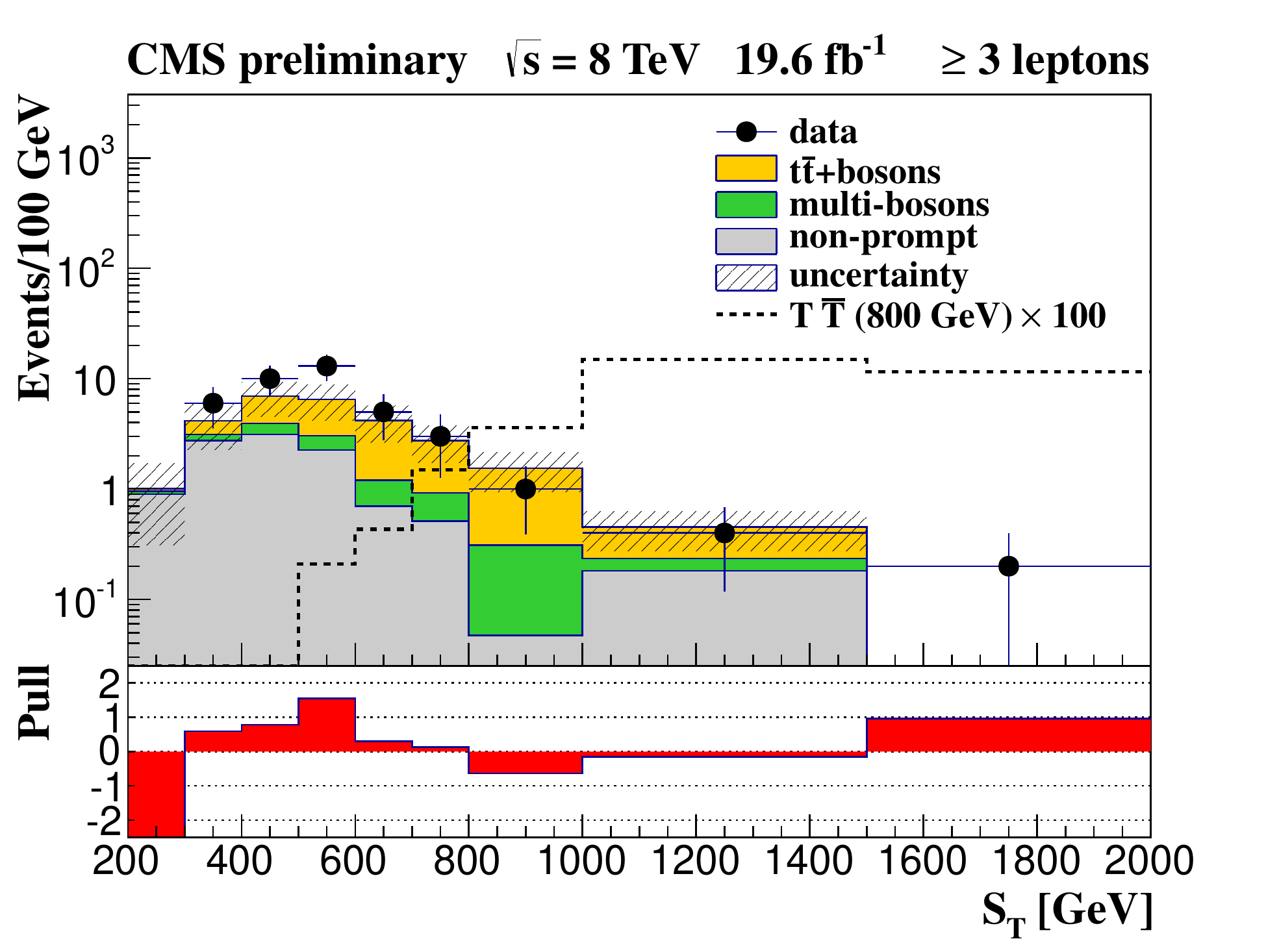}
\caption{Distribution of S$_{T}$ in the same signed and trilepton ($\geq$ 3) categories \label{fig:Multilepton2}}
\end{figure}

\subsection{Results}

\begin{figure}[!h]
\includegraphics[width=0.48\textwidth]{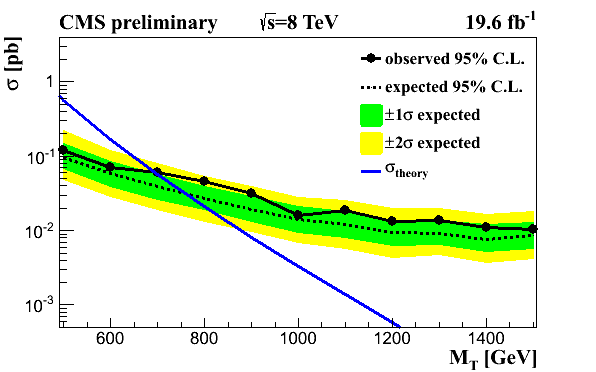}
\includegraphics[width=0.48\textwidth]{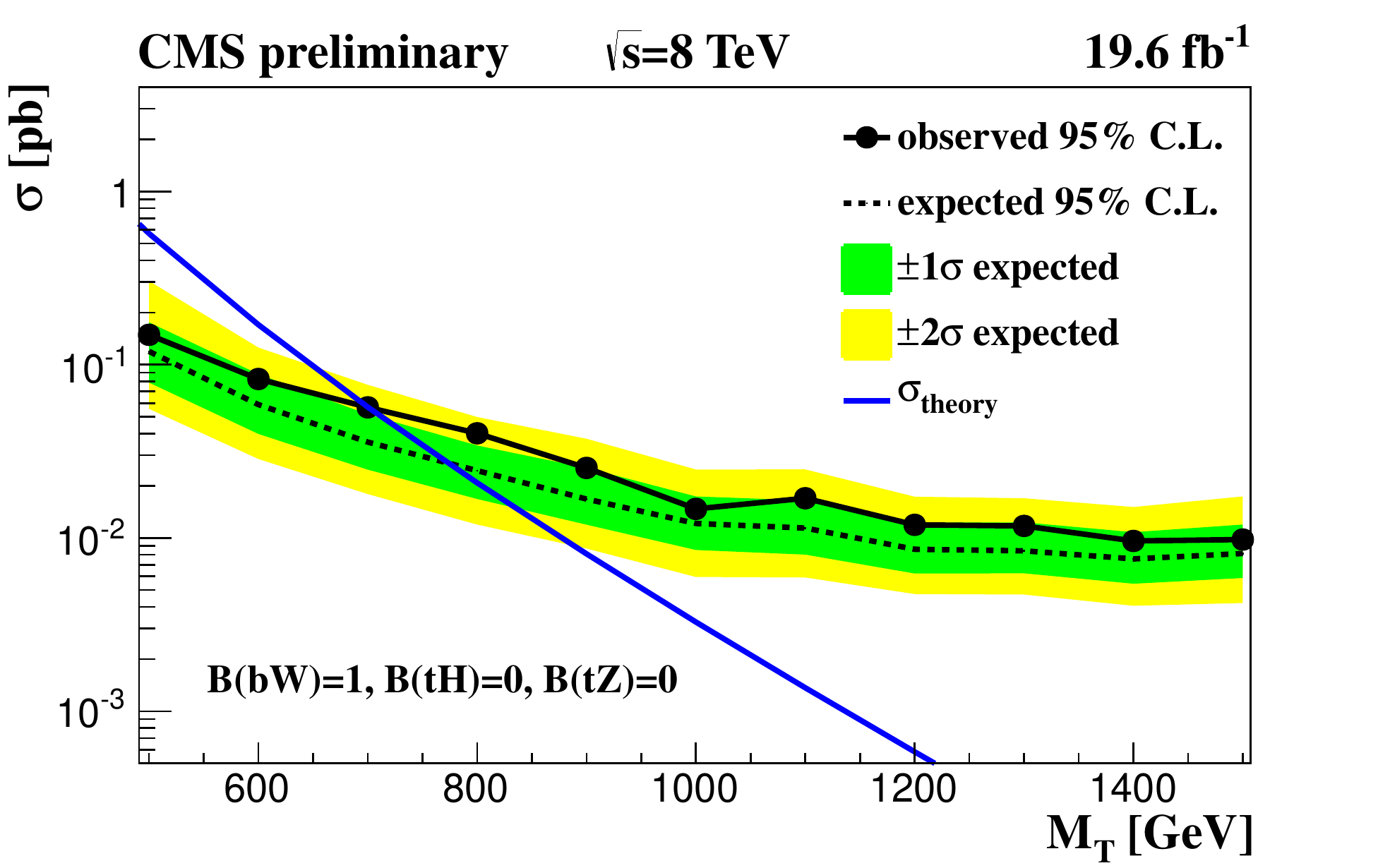}
\caption{Expected and observed limit at 95\% C.L. for T pair production to bW, tH and tZ final states in the nominal (50\% bW, 25\%tZ and 25\%tH) branching ratio scenario (left) and exclusively to bW final state (right) \label{fig:limit1}}
\end{figure}

\begin{figure}[!h]
\includegraphics[width=0.48\textwidth]{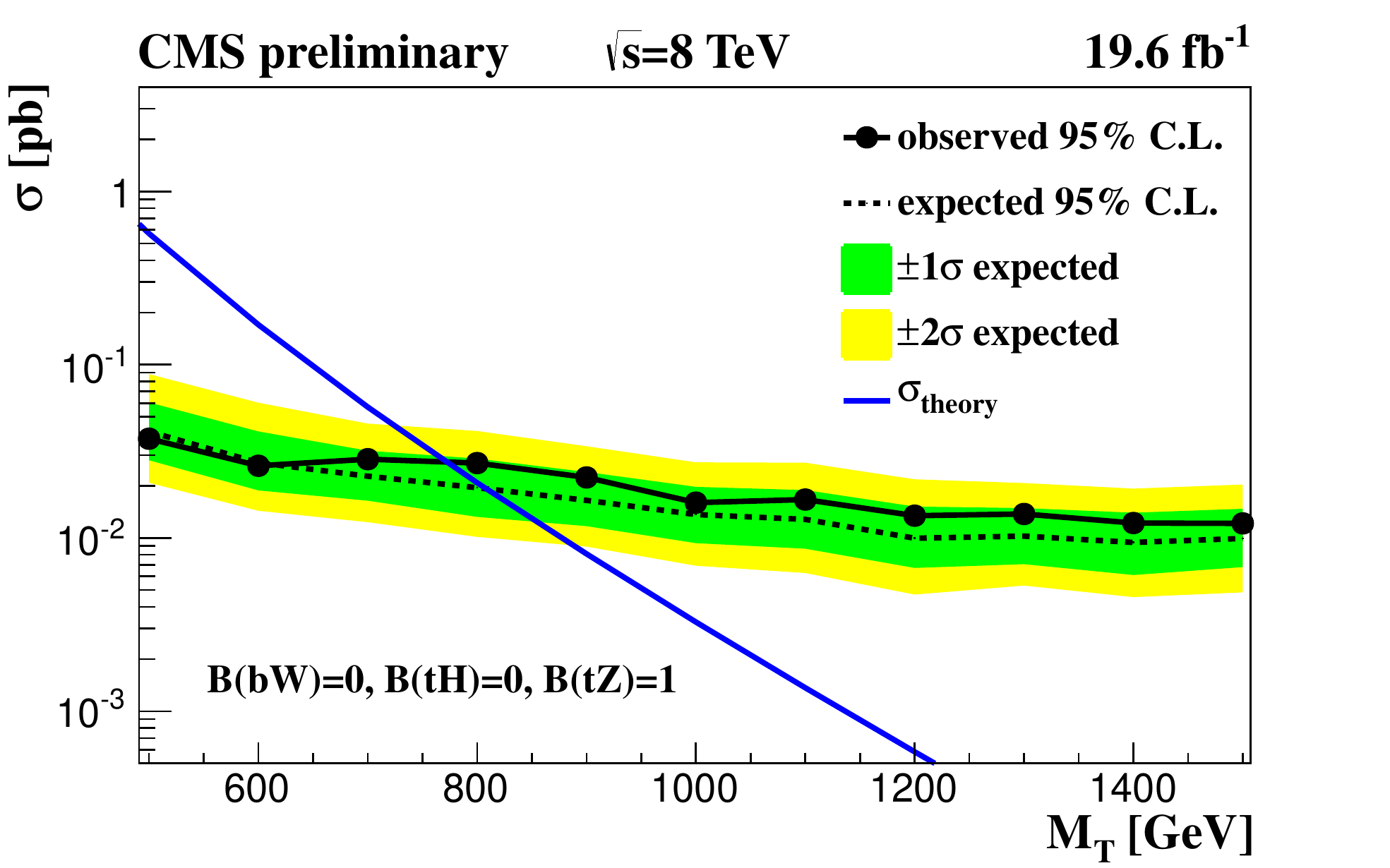}
\includegraphics[width=0.48\textwidth]{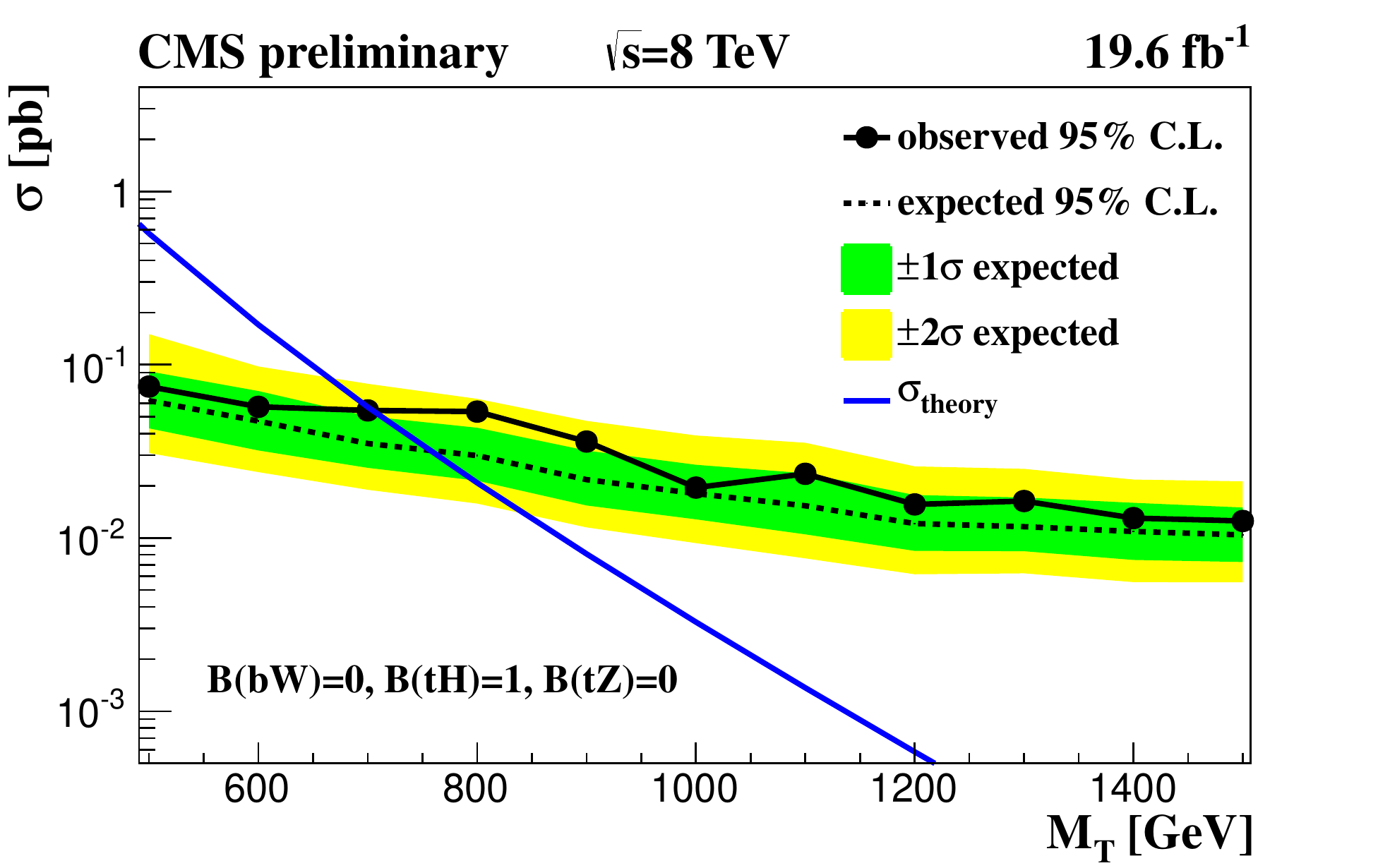}
\caption{Expected and observed limit at 95\% C.L. for T pair production to either exclusively tZ final states (left) or tH final states (right) \label{fig:limit2}}
\end{figure}

\begin{figure}[!h]
\includegraphics[width=0.48\textwidth]{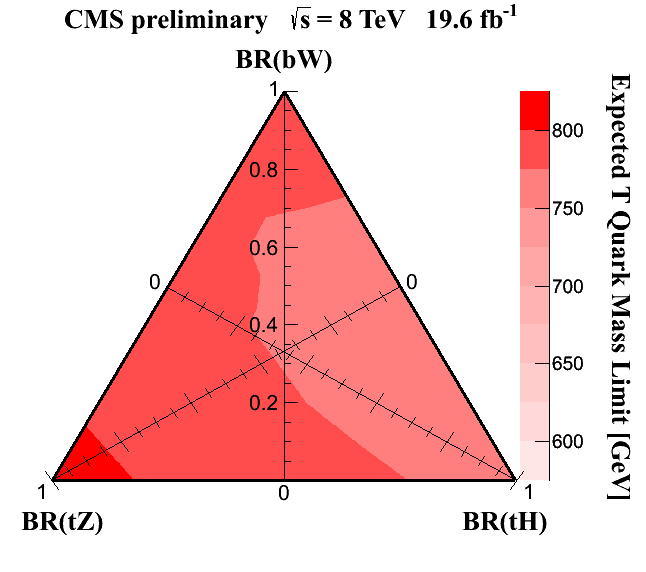}
\includegraphics[width=0.48\textwidth]{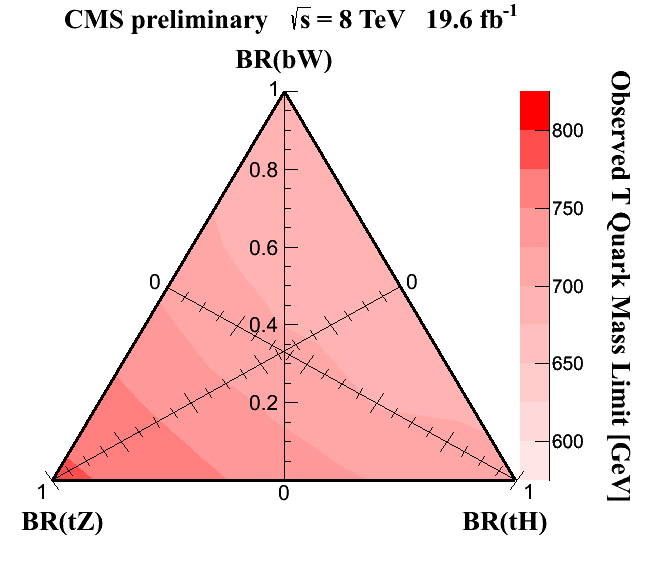}
\caption{Expected and observed limit at 95\% C.L. for T pair production to either exclusively tZ final states (left) or tH final states (right) \label{fig:limit3}}
\end{figure}

In the single lepton final state (Sec.~\ref{sec:Singlelep}), the shape of the BDT distributions is used to obtain the final likelihood using a Bayesian approach ~\cite{theta-stat}. In the mutli-lepton channel (Sec.~\ref{sec:Multilep}), the event yields in 12 bins (corresponding to three dilepton channels, ee, e$\mu$ and $\mu\mu$ and eee, e$\mu \mu$ or ee$\mu$  and $\mu\mu \mu$ in the trilepton case) is used to compute the final likelihood to obtain an exclusion limit. These two channels are then combined to obtain exclusion limits at 95\% C.L. as shown in Fig.{~\ref{fig:limit1}, \ref{fig:limit2}}. Further, we scan across 22 branching fraction scenarios as shown in Fig.{~\ref{fig:limit3}}. Each point on this plot shows the mass exclusion at 95\% C.L. for a given combination of branching ratios. These results represent an inclusive search for the T quark with minimal model dependence.

\clearpage

\section{Search for top partners with charge 5/3 in the same sign dileptonic final state ~\cite{cms_search2}}\label{sec:T53}

\begin{figure}[!h]\center
\includegraphics[width=0.37\textwidth]{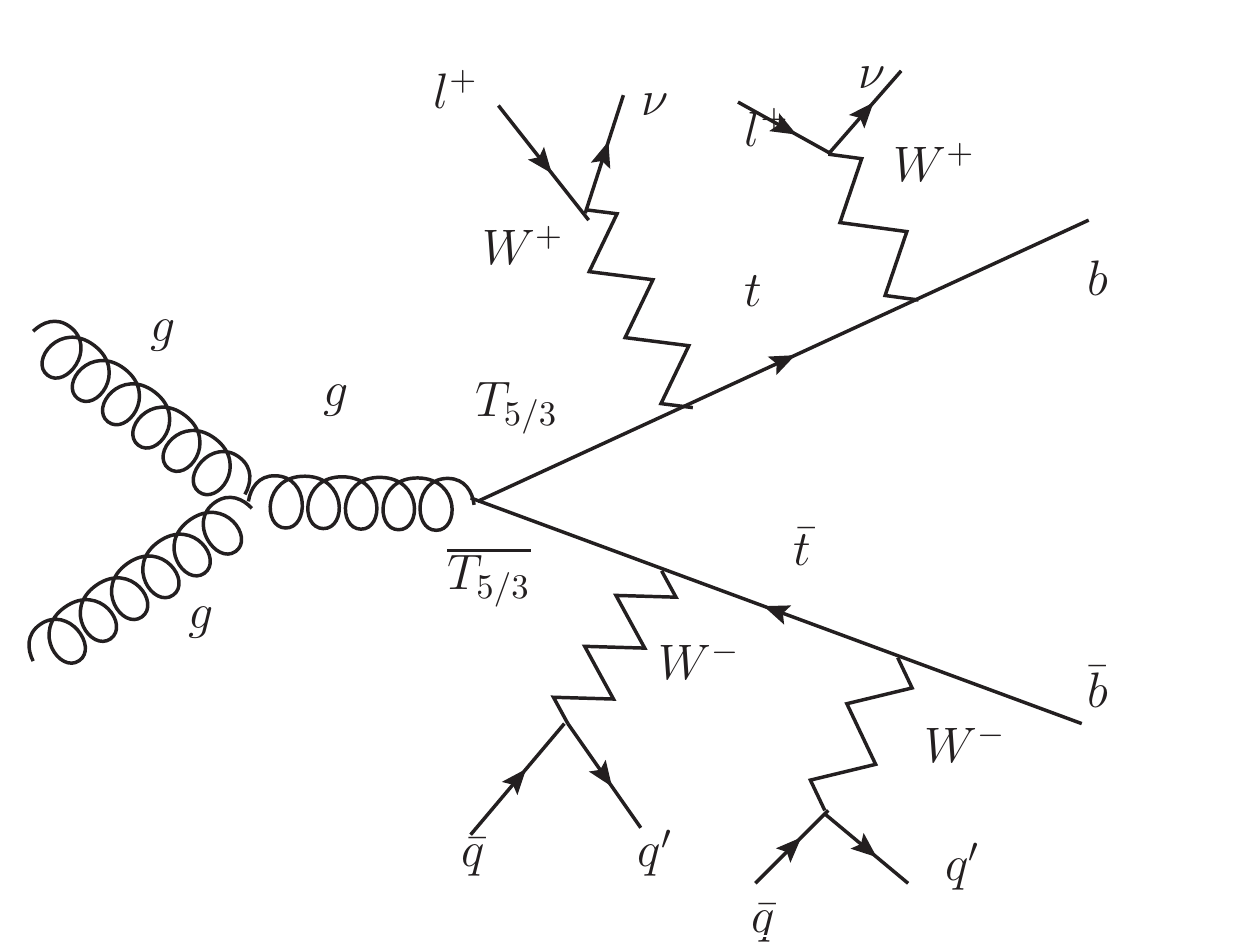}
\includegraphics[width=0.37\textwidth]{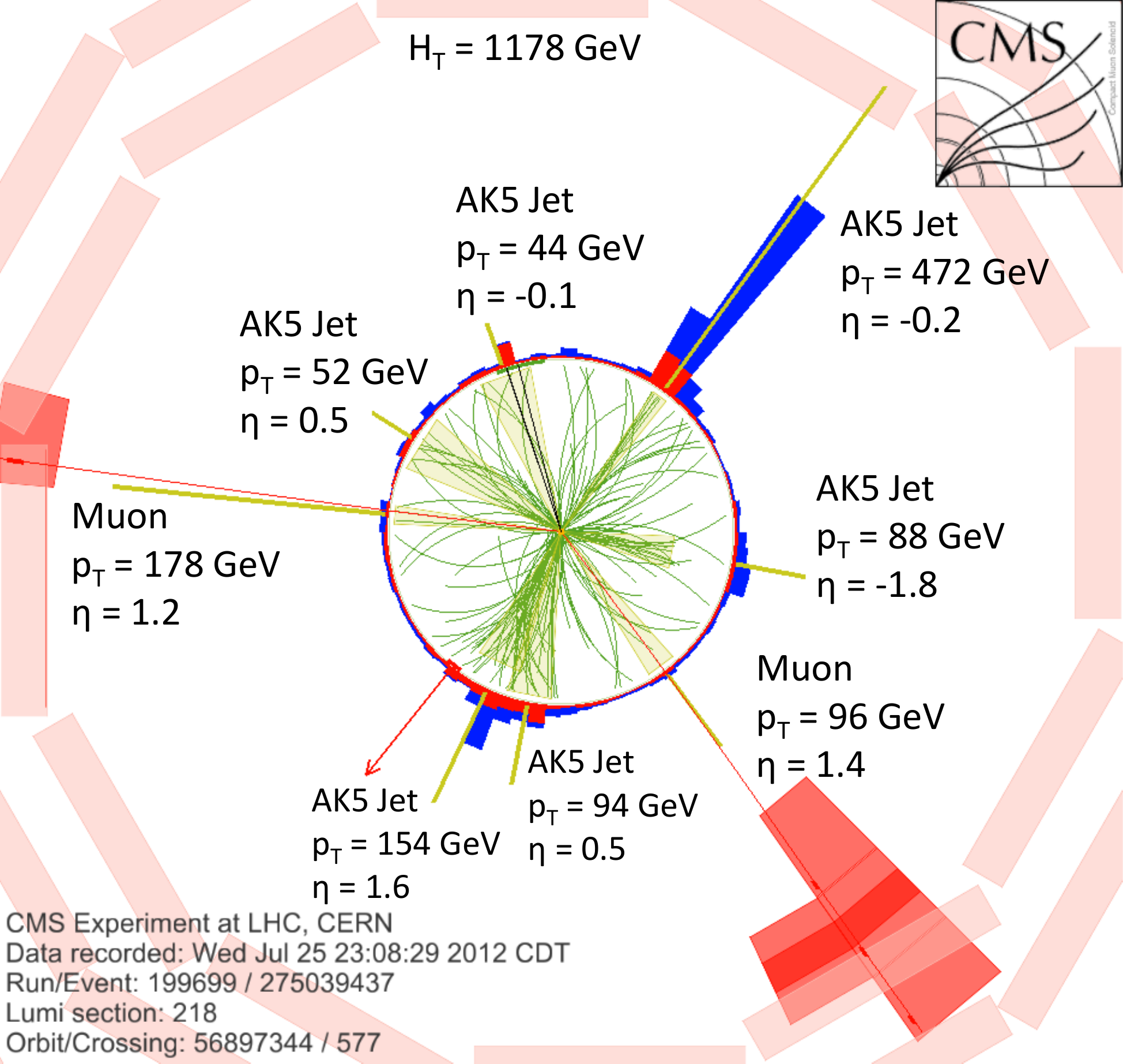}
\caption{Feynman diagram of the process under consideration (left). Candidate event of T$_{5/3}$ pair production (right).\label{fig:T53}}
\end{figure}

A search is carried out for an exotic heavy quark of charge 5/3. The production mechanism along with the decay products is shown in Figure~\ref{fig:T53}. The T$_{5/3} \rightarrow$ bW branching ratio is assumed to be 100\%. The smoking gun signature of these exotic top partners is same-signed dileptons that, should be noted, arise from only one leg of the Feynman diagram. The hadronic W decays can, hence, be utilized to reconstruct the mass of the T$_{5/3}$. Among the various backgrounds, there are SM processes such as diboson and triboson decays. The contribution of these processes to the total background is obtained directly from simulations, since the rate of production of these processes is extremely low and hence are rarely produced in the detector. The instrumental backgrounds such as charge mis-idenification are estimated in a data driven way by estimating the frequency with which one obtains same signed leptonic events within the Z-boson mass window. The backgrounds from fake leptons are obtained from data as described in Section~\ref{sec:Multilep}.

\begin{figure}[!h]\center
\includegraphics[width=0.35\textwidth]{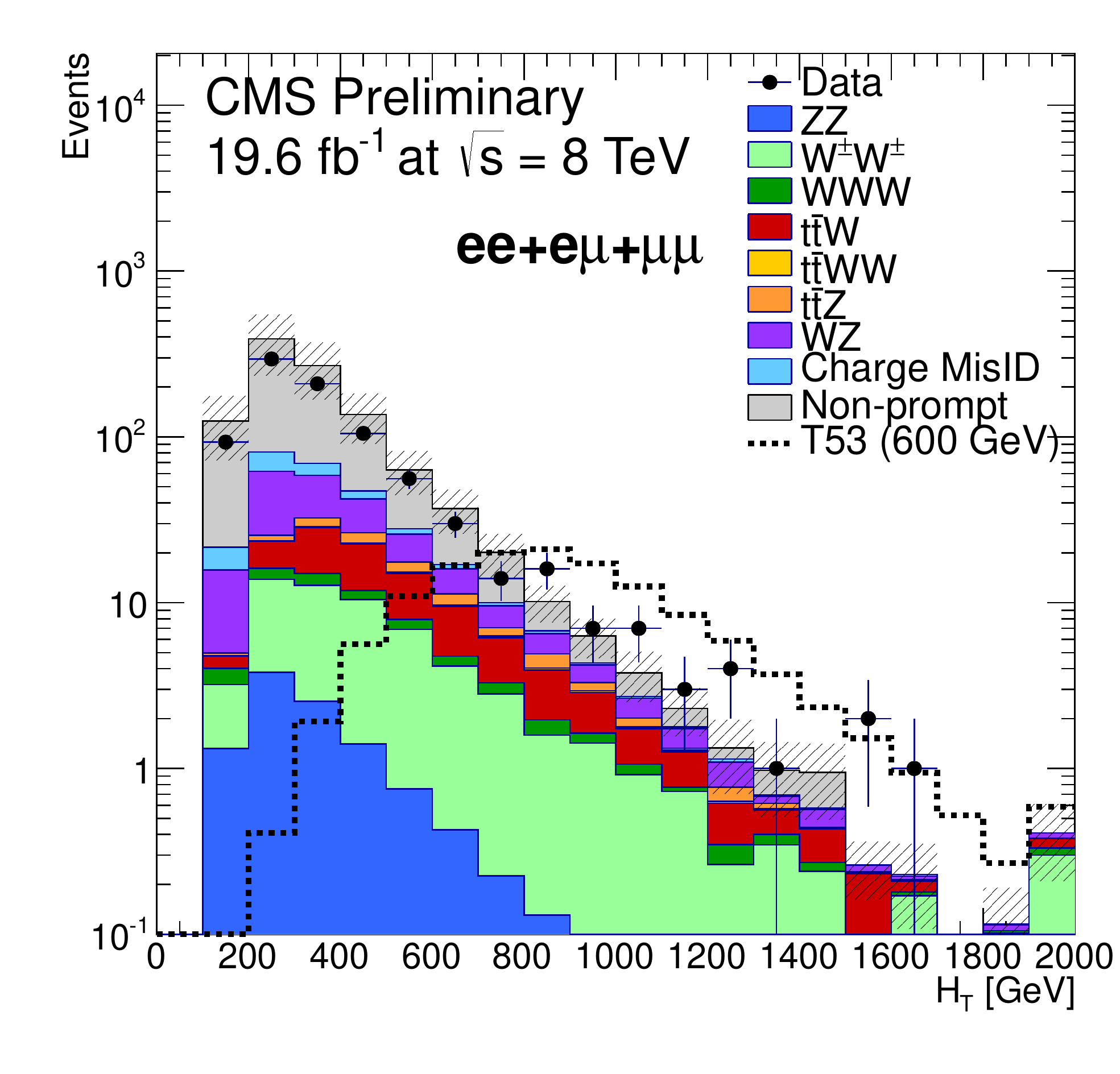}
\includegraphics[width=0.35\textwidth]{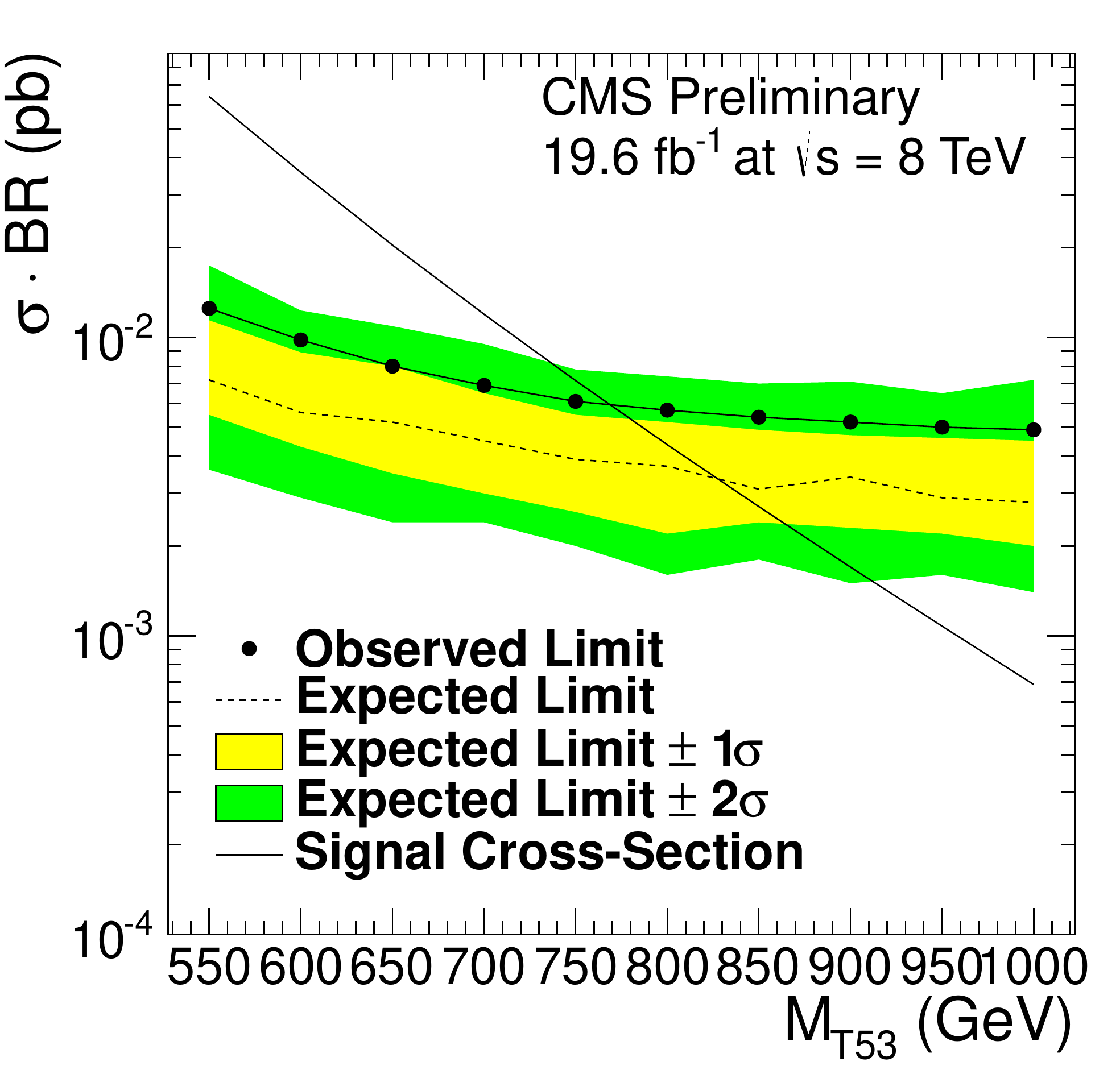}
\caption{H$_{T}$ distribution in the same sign dilepton final state (left).  Exclusion limit obtained at 95\% C.L. (right)\label{HT}}
\end{figure}

The center of mass energy of the LHC allows one to probe T$_{5/3}$ masses where jets from the decay of the heavy quark start merging due to the Lorentz boost experienced by the system of particles. Jet substructure techniques are utilized by constructing the ``jet-constituent'' variable. This variable represents the number of AK5 jets, W-tagged jets weighted by a factor of 2 and top-tagged jets weighted by a factor of 3. The weighting factors are based on the number of hadronic components in W or top-tagged jets. To obtain the final yields (Table~\ref{tab:T53FinalYield}), events that have an invariant mass of the dilepton system within 15 GeV of the Z boson mass window are vetoed, a minimum of five jet constituents are required and H$_{T}$ (shown in Fig~\ref{HT} and computed as the scalar sum of the p$_{T}$ of the jets and the leptons) of the event should be at least 900 GeV.  A candidate T$_{5/3}$ pair production event that passes the event selection requirements is shown in Fig~\ref{fig:T53}. 

\begin{table}[!htbp]
\center

\caption{The total number of expected and observed events after all event selection requirement. The backgrounds include contribution from Monte Carlo driven estimates, the contribution due to fake leptons and from charge mis-identification. The errors indicate systematic as well as statistical errors.}\label{tab:T53FinalYield}
\begin{tabular}{|ccccc|c|}
\hline\hline
         & MC             & Fake leptons  & Charge Mis-ID     & Total Expected  & Observed \\
\hline
$ee$   & 0.7 $\pm$ 0.2 & 1.9 $\pm$ 1.2 & 0.06 $\pm$ 0.02 & 2.6 $\pm$ 1.3 & 0        \\
$e\mu$   & 1.9 $\pm$ 0.4 & 0.6 $\pm$ 0.9 & 0.05 $\pm$ 0.01 & 2.5 $\pm$ 1.0 & 6        \\
$\mu \mu$   & 1.3 $\pm$ 0.3 & 0.2 $\pm$ 0.6 &    -            & 1.5 $\pm$ 0.7 & 5        \\
\hline
All      & 3.9 $\pm$ 0.8 & 2.6 $\pm$ 1.8 & 0.1  $\pm$ 0.02 & 6.6 $\pm$ 2.0 & 11       \\
\hline\hline
\end{tabular}
\end{table}

The final exclusion limits are computed with a frequentist approach, shown in Fig~\ref{HT}.

\subsection{Mass reconstruction}

The reconstruction of the T$_{5/3}$ mass in essential in determining the nature of new physics, if a statistically significant excess is observed in the same-sign dileptonic final state. The mass of the T$_{5/3}$ is reconstructed in the following way:

\begin{itemize}

\item If there exists a CA ÒtopÓ jet (in 22\% signal events), then is it combined with the hadronically decaying ``W'' tagged jet or two AK5 jets with an invariant mass within 20 GeV of m$_{W}$.

\item In the absence of CA ÒtopÓ jets,  two Ws are reconstructed (from CA ÒWÓ jets in 80\% signal events or AK5 jets with an invariant mass within 20 GeV of mW ) and combined with a jet.  The invariant mass of the ÒtopÓ jet is required to be within 30 GeV of the top-quark mass.

\end{itemize}

\begin{figure}[!h]\center
\includegraphics[width=0.30\textwidth]{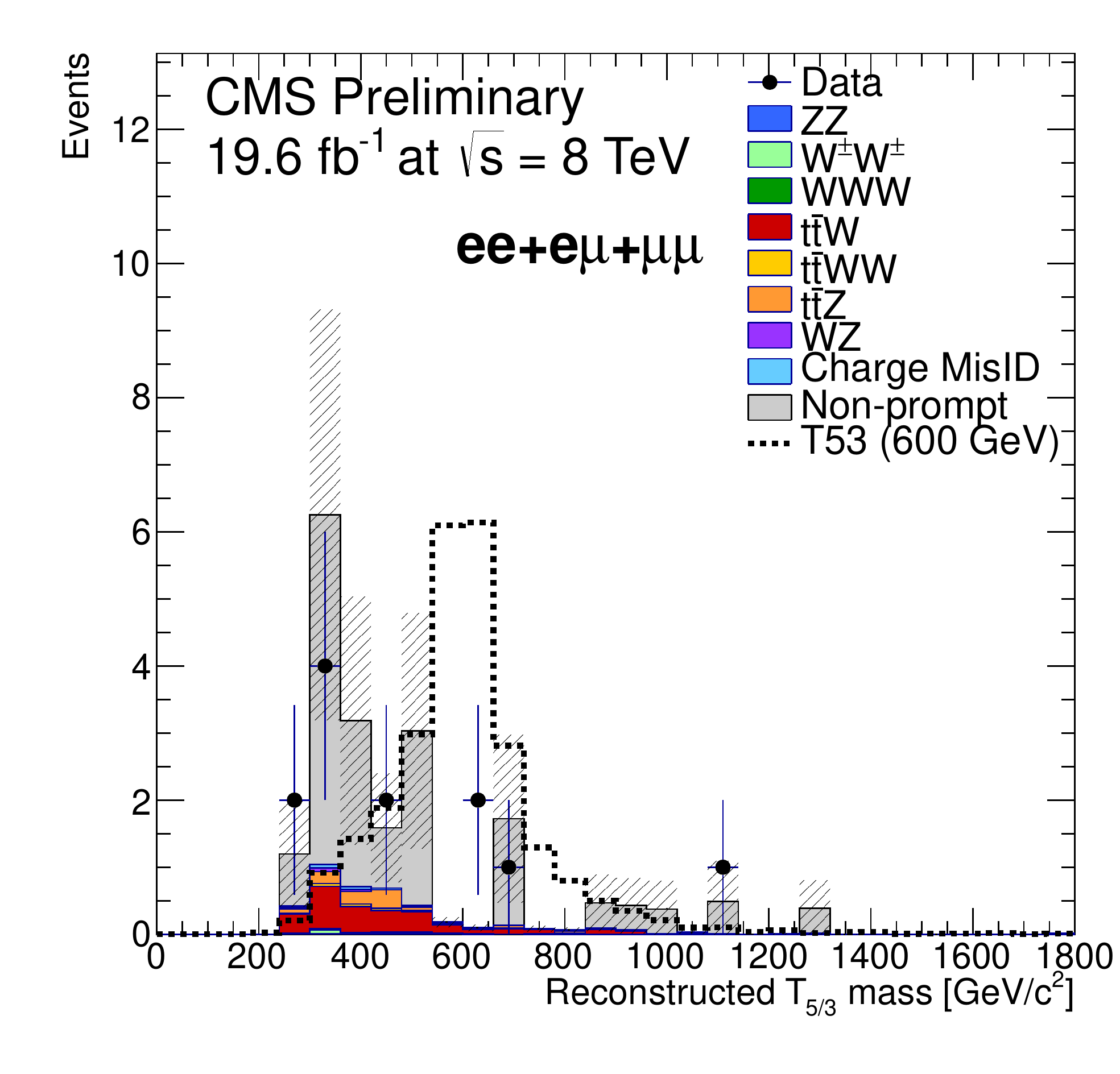}
\caption{Reconstruted T$_{5/3}$ mass.\label{T53Mass}}
\end{figure}

The reconstructed T$_{5/3}$ mass is shown in Fig.~\ref{T53Mass}.

\clearpage

\section{Reinterpretations from RPV SUSY search for b$^{\prime}$ searches ~\cite{cms_search3}}\label{sec:SUSY}

\begin{figure}[!h]
\includegraphics[width=0.48\textwidth]{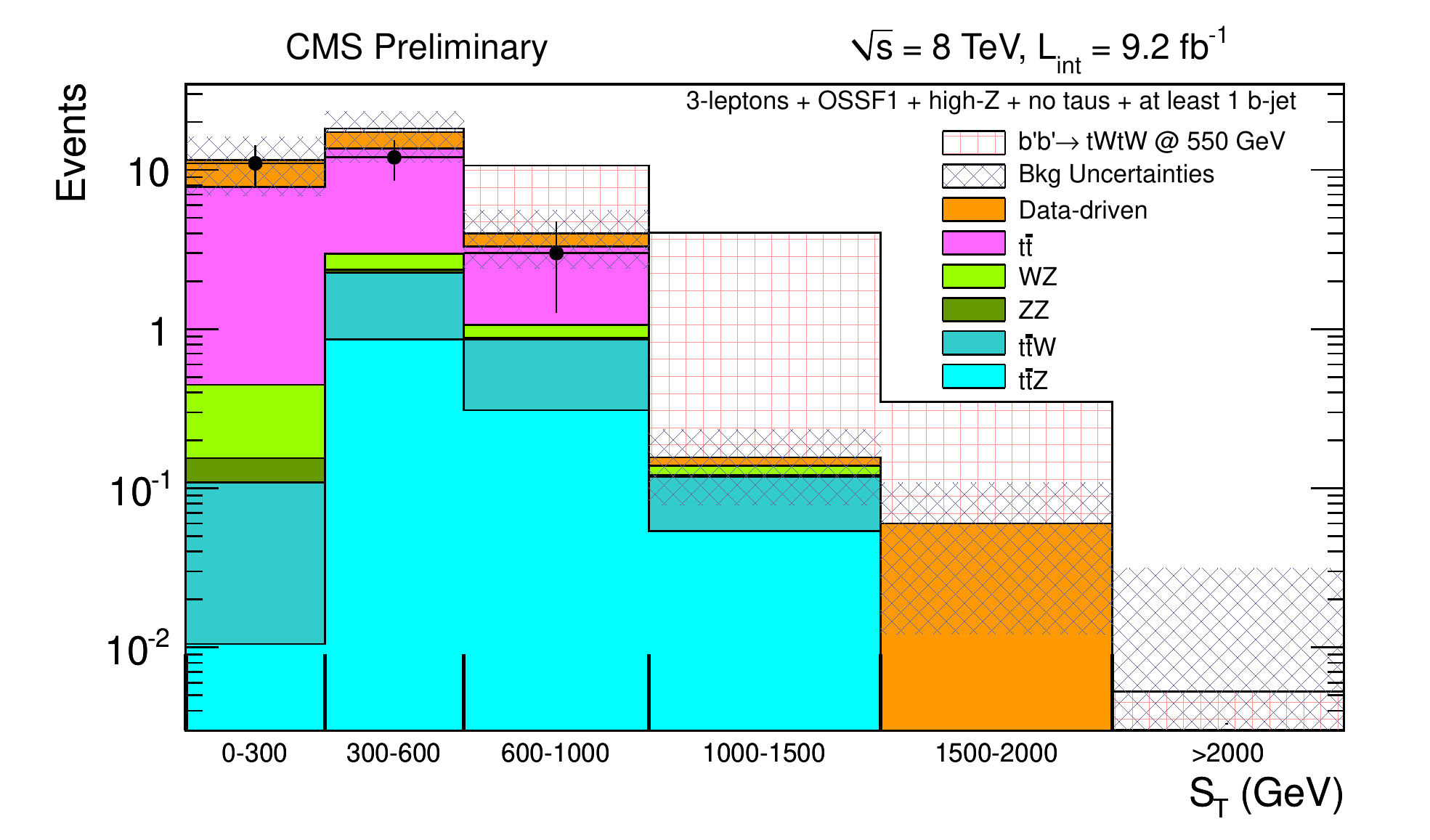}
\includegraphics[width=0.48\textwidth]{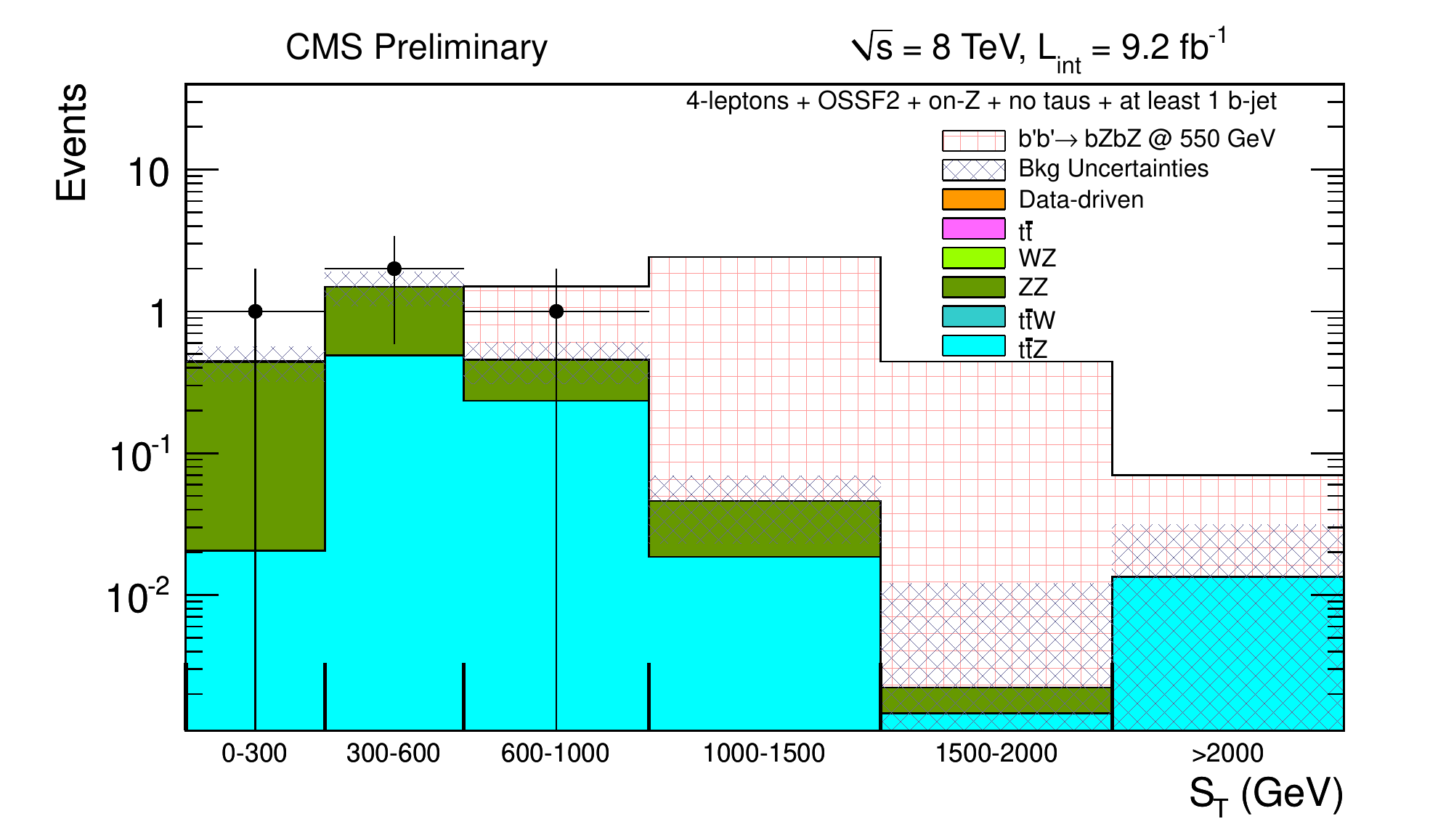}
\caption{S$_{T}$ distribution in the three (left) and four lepton (right) category with requirements on opposite-sign same flavor (OSSF) leptons.\label{fig:SUSb}}
\end{figure}

\begin{figure}[!h]
\includegraphics[width=0.52\textwidth]{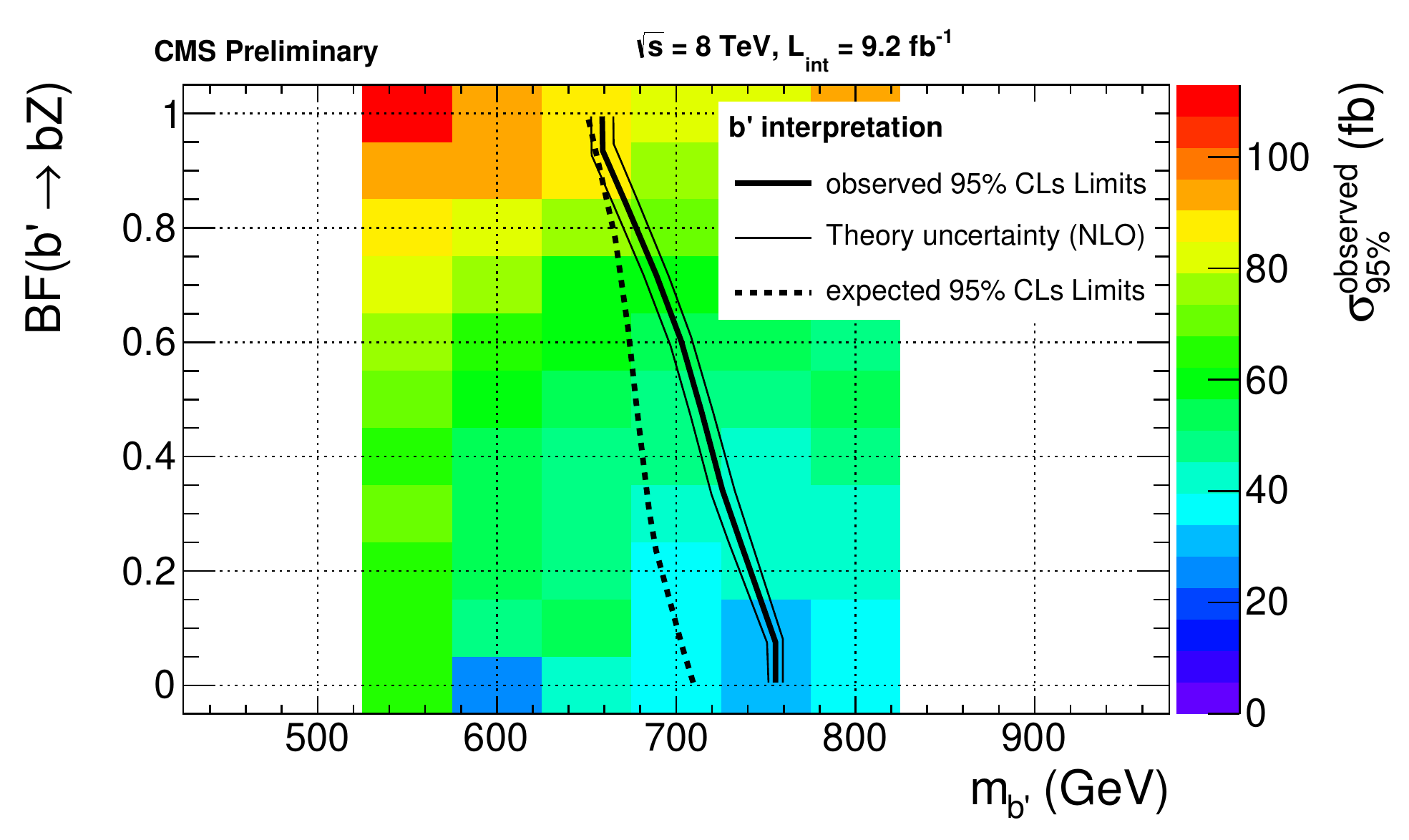}
\includegraphics[width=0.46\textwidth]{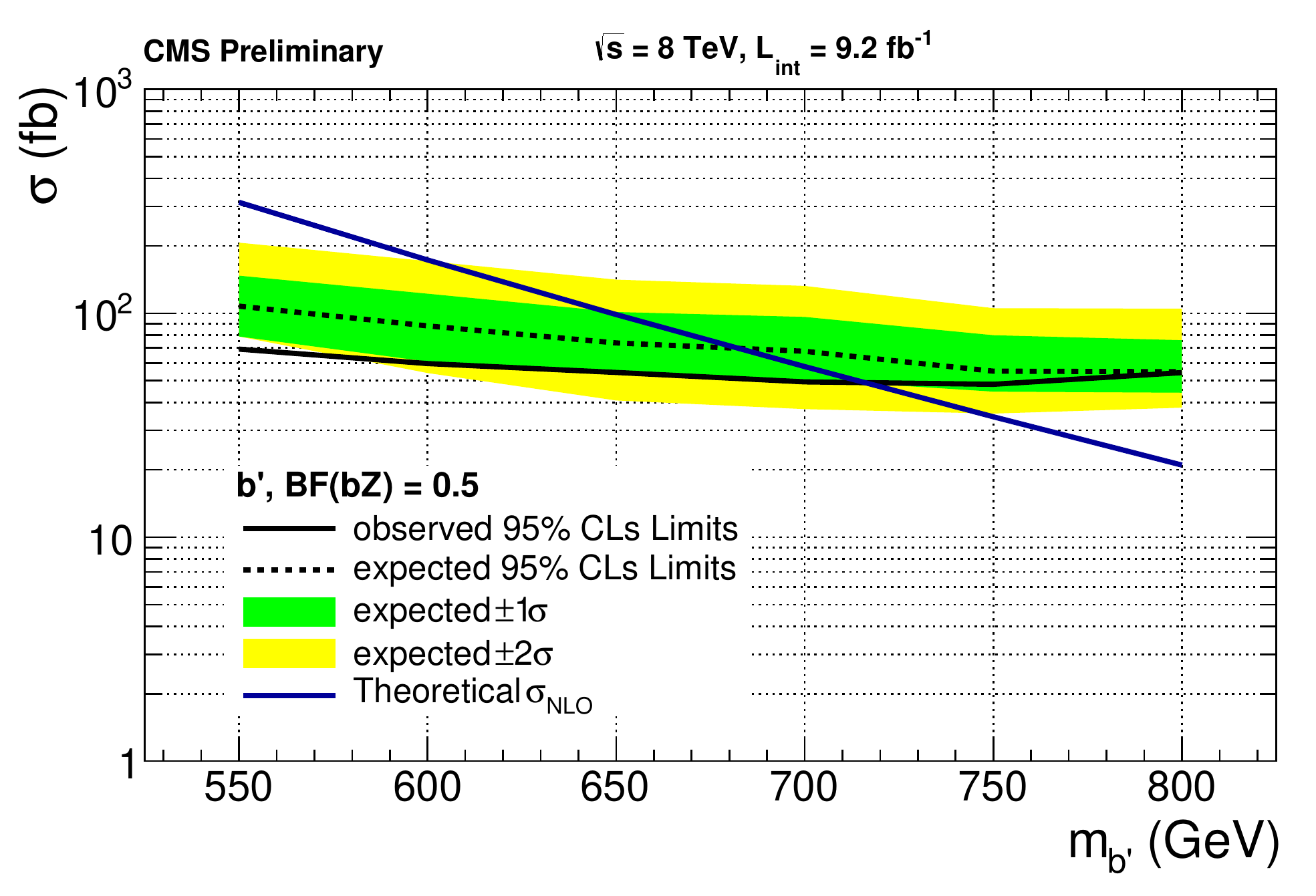}
\caption{Exclusion limit on mass and cross section as a function of the branching ratio (left). The exclusion limit at 95\% C.L. for 50\% branching ratio of b$^{\prime} \rightarrow$ tW and b$^{\prime} \rightarrow$ bZ.\label{fig:SUSf}}
\end{figure}

Dedicated searches for b$^{\prime}$ to tW, bZ and bH have been carried out in CMS. This analysis reinterprets an RPV SUSY search into multilepton final states as a search for b$^{\prime}$. The main discriminating variable under consideration is S$_{T}$. Two distinct event categories are constructed based on the presence of three or four leptons. The final yields are computed in bins of S$_{T}$ as shown in Fig~\ref{fig:SUSb} and the final exclusion limit as a function of branching ratio is shown in Fig.~\ref{fig:SUSf}.

\clearpage

\section{Conclusion}

In the searches quoted above, we do not observe an excess and our findings are consistent with SM predictions. Hence we compute exclusion limits at 95\% C.L. These results are summarized in Table~\ref{tab:summary}.

\begin{table}[!htbp]
\center
\begin{tabular}{|c|c|c|}
\hline\hline
          Model considered             & Branching ratio scenario  & Exclusion Limit \\
\hline
Vector-like T quark & 50\% bW, 25\% tH and 25\% tZ & 696 GeV        \\
Vector-like T quark & 100\%bW & 700 GeV        \\
Vector-like T quark & 100\%tH & 706 GeV        \\
Vector-like T quark & 100\%tZ & 782 GeV        \\
\hline
T$_{5/3}$ & 100\% tW & 770 GeV\\
b$^{\prime}$ &  50\% tW and 50\% bZ & 715 GeV\\
\hline\hline
\end{tabular}
\caption{Summary table of current limits as a function of the branching ratio scenario.\label{tab:summary}}
\end{table}

\end{document}